\documentclass[preprint,showpacs,preprintnumbers,amsmath,amssymb]{revtex4}

\usepackage{graphicx}
\usepackage{dcolumn}
\usepackage{bm}

\begin{document}

\title{Anisotropic Magneto-Thermopower: the Contribution of Interband 
Relaxation} 
\author{J.-E. Wegrowe, Q. Anh Nguyen, M. Al-Barki, J.-F. 
Dayen, T.L. Wade, and H.-J. Drouhin}
\affiliation{Laboratoire des Solides Irradi\'es, Ecole 
Polytechnique, CNRS-UMR 7642 \& CEA/DSM/DRECAM/CAPMAG, 91128 Palaiseau Cedex, France.}

\date{\today} 

\begin{abstract}
Spin injection in metallic normal/ferromagnetic junctions is
investigated taking into account the anisotropic magnetoresistance
occurring in the ferromagnetic layer.  On the basis of a
generalized two-channel model, it is shown that there is an interface resistance
contribution due to anisotropic scattering, besides spin accumulation
and giant magnetoresistance.  The corresponding expression of the
thermoelectric power is derived and compared with the expression 
accounting for
the thermoelectric power produced by the giant magnetoresistance. 
Measurements of anisotropic magneto-thermoelectric power are presented
in electrodeposited Ni nanowires contacted with Ni, Au, and Cu.  It is
shown that this thermoelectric power is generated at the interfaces of
the nanowire.  The results of this study indicate that, while the
giant magnetoresistance and the corresponding thermoelectric power
indicate the role of spin-flip scattering, the observed anisotropic
magneto-thermoelectric power might be the fingertint of interband s-d relaxation
mechanisms.
\end{abstract}

\pacs{72.25.Hg,72.15.If,75.47.De \hfill}

\maketitle


\section{Introduction}

In order to explain the high resistance and the high thermoelectric
power observed in transition metals, Mott introduced the concept of
spin-polarized current and suggested that s-d interband scattering
plays an essential role in the conduction properties \cite{Mott}.  This
approach, in terms of two conduction bands, explained the existence of
a spin-polarized current in the 3d ferromagnetic materials, and was used
for the description of anisotropic magnetoresistance (AMR)
\cite{Potter0,Potter}, and thermoelectric power \cite{Handbook}.  With
the discovery of giant magnetoresistance (GMR) \cite{GMR} and related
effects, the development of spintronics focused the discussion on
spin-flip scattering occurring between spin-polarized conducting
channels.  The two-channel model, which describes the
conduction
electrons with majority and minority spins, is applied with great
efficiency to GMR and spin injection effects
\cite{Johnson,VanSon,Valet,Zhang,PRBThermo}, including
metal/semiconductor \cite{Molenkamp} and metal/supraconductor
interfaces \cite{JedemaSupra}.  In this context, it is sufficient to
describe the diffusion process in terms of spin-flip
scattering without the need to invoke interband s-d scattering.

Magneto-thermoelectric power 
(MTEP) experiments in GMR
structures \cite{MTEP,piraux,Shi,Shi2,Tsymbal,Gravier,Gravier2}
however point out the need for a deeper understanding of the dissipative
mechanism responsible for the giant magnetothermopower related to
GMR. The problem of s-d electronic relaxation at the interface was
also put forward in the context of current induced magnetization
reversal mechanisms in various systems exhibiting AMR
\cite{BergerDW,Derek,JulieDW,Marcel,cond-mat}.  However, the interface
contribution to the resistance in relation to AMR has so far not been
investigated.  The aim of the present work is to study the
non-equilibrium contribution of a normal/ferromagnetic (N/F) interface
to both the resistance and the thermoelectric power.

For our purpose, it is convenient to generalize the two spin channel approach to any
relevant transport channels, i.e. to any distinguishable electron
populations $\alpha$ and $\gamma$ \cite{hybride}.  The local out-of-equilibrium state
near the junction is then described by a non-vanishing
chemical-potential difference between these two populations: $\Delta
\mu_{\alpha \gamma}¥ = \mu_{\alpha}-\mu_{\gamma} \neq 0$
\cite{PRBThermo}.  Corollarilly, assuming that the presence of a
junction induces a deviation from the local equilibrium, the $\alpha$
and $\gamma$ populations can be {\it defined by the $\alpha
\rightarrow \gamma$ relaxation mechanism} itself, that allows the
local equilibrium to be recovered in the bulk material ($lim_{z
\rightarrow \pm \infty}¥\Delta \mu(z) = 0$).  In this context
\cite{PRBThermo,cond-mat}, the basic idea we develop here is that,
beyond spin-flip relaxation, interband s-d relaxation also plays a
crucial role in the interface magnetoresistance of magnetic
nanostructures.  Though similar ideas have been suggested in previous
spintronics studies \cite{Mott,Potter0,Potter,Suzuki,Tsymbal,Baxter},
the originality of this work is to deal with interband
relaxation on an equal footing with spin-flip relaxation
\cite{cond-mat} in the framework of a {\it thermokinetic approach}. 
For this purpose, the two spin-channel model is generalized, with the
introduction of the corresponding transport coefficients: the
conductivities $\sigma_{\alpha}$ and $\sigma_{\gamma}$ of each channel
define the total conductivity $\sigma_{t}=¥\sigma_{\alpha} +
\sigma_{\gamma}$ and the conductivity asymmetry $\beta =
(\sigma_{\alpha}- \sigma_{\gamma}) /\sigma_{t}$; the relaxation
between both channels is described by the parameter $L$ (or
equivalently, the relevant relaxation times $\tau_{\gamma
\leftrightarrow \alpha}$).  It is shown that this two-channel model
can be applied straightforwardly to the description of MTEP, by
introducing an extra transport parameter which is nothing but the
derivative of $\beta$ with respect to the energy.  The predictions of
the model are compared with experimental results of anisotropic
MTEP measured in electrodeposited nanowires.

The article is structured as follows: General expressions of the
interface contributions of resistance (Sec.  II) and thermoelectric power (Sec. 
III) are derived, and applied to the case of AMR and GMR systems (Sec. 
IV), and to the corresponding MTEP (Sec.  V).  It is shown that a
contribution of the interface resistance related to AMR and
the corresponding MTEP should be expected.  The experimental study
performed on single-contacted Ni nanowires (Sec.  VI) confirms the
presence of an anisotropic MTEP, which is produced by the interfaces.

\section{Out-of-equilibrium resistance}

In the framework of the two conducting-channel model, which includes
relaxation from one channel to the other, it is possible to show, on
the basis of the entropy variation \cite{PRBThermo}, that the kinetics
are described by the following Onsager equations:

\begin{equation}
\begin{array} {lll}
J_{\alpha} = -\frac{\sigma_{\alpha }}{e} \frac{\partial 
\mu_{\alpha}}{\partial z}\\ 
J_{\gamma} = -\frac{\sigma_{\gamma}}{e} \frac{\partial \mu_{\gamma}}{\partial z}\\ 
\dot{\Psi}_{\alpha \gamma}¥ = L \left ( \mu_{\alpha}-\mu_{\gamma} \right )
\end{array}
\label{Onsager0}
\end{equation}

Where $\dot{\Psi}_{\alpha \gamma}$ describes the relaxation from the channel $\alpha$
to the other channel $\gamma$ in terms of velocity of the reaction $\alpha
\rightarrow \gamma$.  The Onsager coefficient $L$ is
inversely proportional to the relaxation times $\tau_{\alpha 
\leftrightarrow \gamma}$ :

\begin{equation}
L \propto \left ( \frac{1}{\tau_{\alpha \rightarrow
\gamma}} + \frac{1}{\tau_{\gamma \rightarrow \alpha}}\right )
	\end{equation}
	
The out-of-equilibrium configuration is quantified by
the chemical affinity $\Delta \mu = \mu_{\alpha} - \mu_{\gamma}$, 
i.e. the
chemical potential difference of the reaction.

Furthermore, in the case of a stationary regime, the conservation 
laws lead to :

\begin{equation}
   \begin{array} {ll}
\frac{d J_{\alpha}}{dt}\, = \,-\frac{\partial J_{\alpha}}{\partial 
z} - \, \dot{\Psi} = 0 \\ 
\frac{dJ_{\gamma}}{dt}\, = \,-\frac{\partial J_{\gamma}}{\partial z} + \, \dot{\Psi} = 0\\
\end{array} 
\label{con}
\end{equation}

The total current $J_{t}$ is constant:
  
 \begin{equation}
J_{t} = J_{\alpha} + J_{\gamma} = -\frac{1}{e} \frac{\partial 
}{\partial z} \left (\sigma_{\alpha } \mu_{\alpha}+ 
\sigma_{\gamma } \mu_{\gamma } \right ) 
\end{equation}
 
The expression of Ohm's law, $J_{t}= -\sigma_{t} \frac{\partial \Phi}{\partial 
z}$,
is recovered by introducing the measured electric potential $\Phi$ and the
total conductivity $\sigma_{t}=\sigma_{\alpha}+ \sigma_{\gamma}$ 
\cite{Constantes} :

\begin{equation}
e \Phi= \frac{1}{\sigma_{t}}( \sigma_{\alpha} \mu_{\alpha} + 
\sigma_{\gamma} \mu_{\gamma})
\end{equation}

Let us assume that the two channels collapse to a unique conduction
channel for a specific configuration, the reference, which is a
local equilibrium situation: $\Delta \mu_{eq}=0$.  The
out-of-equilibrium contribution to the resistance, $R^{ne}$, is
calculated through the relation:

\begin{equation}
-J_{t}e \, R^{ne}  = \int_{A}^{B}  \frac{\partial }{ \partial z}
(\mu_{\alpha} - e \Phi(z))dz = \int_{A}^{B}  \frac{\partial }{ \partial z}
(\mu_{\gamma} - e \Phi(z))dz 
\end{equation}

so that 

\begin{equation}
 R^{ne}= -\frac{1}{J_{t}e} \int_{A}^{B} 
\frac{\sigma_{\alpha} - \sigma_{\gamma}}{2 \sigma_{t}} \frac{\partial 
\Delta \mu}{ \partial z}dz 
\label{Res}
\end{equation}

where the measurement points $A$ and $B$ are located far enough from
the interface (inside the bulk) so that $\Delta 
\mu (A)=\Delta \mu (B) =0$.  The derivative is only calculated in 
the intervals where
$\Phi$ is continuous.  The above relation allows the
out-of-equilibrium resistance at a simple junction between two layers
(composed by the layers $I$ and $II$) to be easily calculated.  If the
junction is set at $z=0$ and the conductivities are respectively
$\sigma_{i}^{I}$ and $\sigma_{i}^{II}$ ($i=\{\alpha, \gamma \}$), we
have:

\begin{equation}
-J_{T}e \, R^{ne} = \int_{A}^{0} 
\frac{\sigma^{I}_{\alpha} - \sigma^{I}_{\gamma}}{2 \sigma_{T}} \frac{\partial 
\Delta \mu^{I}}{ \partial z}dz + \int_{0}^{B} 
\frac{\sigma^{II}_{\alpha} - \sigma^{II}_{\gamma}}{2 \sigma_{T}} \frac{\partial 
\Delta \mu^{II}}{ \partial z}dz
\label{ResJunct}
\end{equation}
The equilibrium is recovered in the bulk, so that:

\begin{equation}
R^{ne} = \left ( 
\frac{\sigma^{I}_{\alpha} - \sigma^{I}_{\gamma}}{\sigma^{I}_{t}} -
\frac{\sigma^{II}_{\alpha} - \sigma^{II}_{\gamma}}{\sigma^{II}_{t}} 
\right ) 
\frac{\Delta \mu(0)}{2 J_{t}e}
\label{Result}
\end{equation}

The chemical potential difference $\Delta \mu(z)$, which accounts for
the pumping force opposed to the relaxation $\alpha \rightarrow 
\gamma$, is obtained by solving the diffusion equation deduced
from Eqs.  (\ref{Onsager0}) and (\ref{con}) 
\cite{Johnson,VanSon,Valet,Zhang,PRBThermo}:

\begin{equation}
\frac{\partial^{2}¥\Delta \mu(z)}{\partial z^{2}}= \frac{\Delta \mu(z)}{l_{diff}^{2}}  
\label{DiffEq}
\end{equation}

where 
\begin{equation}
l_{diff}^{-2}= eL(\sigma_{\alpha}^{-1}+\sigma_{\gamma}^{-1})
\label{ldiff}
\end{equation}

is the diffusion length related to the $\alpha \rightarrow \gamma$ relaxation.

At the interface ($z=0$), the continuity of the currents for each channel 
writes:
\begin{equation}
J_{\alpha}(0)=-\frac{\sigma_{\alpha} \sigma_{\gamma}}{e 
\sigma_{t}} \frac{\partial \Delta \mu}{\partial 
z}+\frac{\sigma_{\alpha}}{\sigma_{t}} J_{t}=J_{\gamma}(0)
\label{currentcon}
\end{equation}

 which leads to the general relation: 
 
\begin{equation}
 \Delta \mu (0)= \left ( \frac{\sigma^{I}_{\alpha}}{\sigma^{I}_{t}} - 
\frac{\sigma^{II}_{\alpha}}{\sigma^{II}_{t}} \right 
) \, \left ( \frac{\sigma^{I}_{\alpha} \sigma^{I}_{\gamma}}{\sigma^{I}_{t} l_{diff}}+ 
\frac{\sigma^{II}_{\alpha}\sigma^{II}_{\gamma}}{\sigma^{II}_{t} 
l^{II}_{diff}} \right )^{-1}\, \, eJ_{T}
\label{DeltaMu0}
\end{equation}

Inserting Eq. (\ref{DeltaMu0}) into Eq.  (\ref{Result}), we obtain the general expression for
the out-of-equilibrium resistance (per unit area) produced by the $\alpha \rightarrow
\gamma$ relaxation mechanism 
at a junction:

\begin{equation}
R^{ne} = \left ( 
\frac{\sigma^{I}_{\alpha} - \sigma^{I}_{\gamma}}{2 \sigma^{I}_{t}} -
\frac{\sigma^{II}_{\alpha} - \sigma^{II}_{\gamma}}{2 \sigma^{II}_{t}} 
\right )\, \left ( \frac{\sigma^{I}_{\alpha}}{\sigma^{I}_{t}} - 
\frac{\sigma^{II}_{\alpha}}{\sigma^{II}_{t}} \right 
) \, \left ( \sqrt{\frac{\sigma^{I}_{\alpha} 
\sigma^{I}_{\gamma}eL^{I}}{\sigma^{I}_{t}}} + \sqrt{\frac{\sigma^{II}_{\alpha} 
\sigma^{II}_{\gamma}eL^{II}}{\sigma^{II}_{t}}}\right )^{-1}
\label{Rout}
\end{equation}

It is convenient to describe the conductivity asymmetry by a 
parameter $\beta$ such that $\sigma_{\alpha}= \sigma_{t} (1+\beta)/2$ and
$\sigma_{\gamma}=\sigma_{t}(1-\beta)/2$. The out-of-equilibrium 
contribution to the resistance then takes the following form:

\begin{equation}
R^{ne} = \frac{1}{2}
\frac{(\beta_{I} - \beta_{II})^{2}}{\sqrt{eL^{I}\sigma_{t}^{I}(1-\beta_{I}^{2})}+ 
\sqrt{eL^{II}\sigma_{t}^{II}(1-\beta_{II}^{2})}}
\label{Rbeta}
\end{equation}

where the diffusion length $l_{diff}$ now writes:

\begin{equation}
	l_{diff}^{-1}= 2 \sqrt{\frac{eL}{\sigma_{t}¥(1-\beta^{2})}}
	\end{equation}

\section{giant magnetoresistance vs. anisotropic magnetoresistance}

\subsection{Giant Magnetoresistance}

The most famous example of the out-of-equilibrium resistance described
in the preceding section, is the giant magnetoresistance (GMR)
\cite{GMR} occurring near a junction composed of two ferromagnetic
layers $F_{1}/F_{2}$ made out of the same metal.  The electronic populations are the
spin-polarized carriers quantized along the ferromagnetic order
parameter $\alpha = \uparrow$, $\gamma= \downarrow $.  The diffusion
length is the spin-diffusion length $l_{diff}=l_{sf}$.  The $\alpha
\rightarrow \gamma$ relaxation is the spin-flip relaxation, and tends
to balance the deviation from the local equilibrium.  This process
leads to a spin-accumulation described by the generalized
force $\Delta \mu = \mu_{\uparrow}-\mu_{\downarrow}$.  The local
equilibrium ($\Delta \mu = 0 $) is recovered in the bulk ferromagnet,
at the voltage probes, or equivalently in the case of two parallel
magnetic configurations.  When the magnetization of the two layers are
parallel, we have indeed:
$\sigma_{\uparrow}^{I}=\sigma_{\uparrow}^{II}$ and
$\sigma_{\downarrow}^{I}=\sigma_{\downarrow}^{II}$, and $R^{ne}=0$. 
In contrast, for an
antiparallel configuration
$\sigma_{\uparrow}^{I}=\sigma_{\downarrow}^{II}$ and
$\sigma_{\downarrow}^{I}=\sigma_{\uparrow}^{II}$.  In terms of
conductivity asymmetry $\beta_{s}$, we have
$\sigma_{\uparrow}= \sigma_{t} (1+\beta_{s})/2$ and
$\sigma_{\downarrow}=\sigma_{t}(1-\beta_{s})/2$ (the subscript $s$
refers to the $s$ type - possibly $s d$ hybridized - conduction band). 
The out-of-equilibrium resistance writes:

\begin{equation}
R_{GMR}^{\uparrow \downarrow } =  \frac{
\beta_{s}^{2}}{\sigma_{t}¥(1-\beta_{s}^{2})} \, l_{sf} =
 \frac{\beta_{s}^{2}}{\sqrt{eL\sigma_{t}(1-\beta_{s}^{2})}}
 \label{RGMR}
\end{equation}

This expression is the well-known giant magnetoresistance
\cite{Johnson,VanSon,Valet,Zhang,PRBThermo,Jedema,George} measured in
various $F_{1}¥/N/F_{2}$ devices.  It is usually presented as the
normalized ratio 

\begin{equation}
\frac{R_{GMR}^{\uparrow \downarrow }}{R_{0}} =
\frac{\beta_{s}^{2}}{1-\beta_{s}^{2}} \, \frac{l_{sf}}{\Lambda}
\label{GMRStandard}
\end{equation}
measured on a layer of thickness $\Lambda$, where $R_{0} = R^{\uparrow
\uparrow} = R^{\downarrow \downarrow} = \Lambda / \sigma_{t}$ is the overall
resistance of the layers (also per surface units). 

In the case of a single $N/F$ junction, we have
$\sigma_{\alpha}^{I}=\sigma_{\gamma}^{I}$ in the normal metal and
$\sigma_{\alpha}^{II} \neq \sigma_{\gamma}^{II}$ in the ferromagnetic
metal.  The out of equilibrium resistance writes:

\begin{equation}
R_{GMR}^{N-F} = \frac{1}{2}
\frac{\beta^{2}_{s}}{\sqrt{eL^{N}\sigma^{N}_{t}}+\sqrt{eL^{F}\sigma^{F}_{t}(1-\beta^{2}_{s})}} 
\end{equation}

This is the out-of-equilibrium resistance arising in a single
magnetic layer.  It is worth pointing out that, in spite of the
existence of spin accumulation and non-vanishing out-of-equilibrium
resistance, it is not possible to measure a deviation of
$R_{GMR}^{N-F}$ from a reference state because the resistance does not
vary with the magnetic configurations, or with any well-controled external
parameters (except in the case of domain wall
scattering, discussed e.g. in reference \cite{DWS}).  In other words,
$R_{GMR}$ is present but there is nevertheless no analyzer, or probe,
to detect it.  Although the GMR results are well known, the more
general Eq.  (\ref{Rout}) allows one to push the discussion about
non-equilibrium resistances beyond GMR effects.

\subsection{Out-of-equilibrium anisotropic magnetoresistance}

 From our generalized approach one should predict the existence of a
 {\bf non-equilibrium anisotropic magnetoresistance} (NeAMR).  The
 anisotropic magnetoresistance (AMR) is characterized by a
 conductivity $\sigma_{t}(\theta)$ which depends on the angle $\theta
 = (\vec{I},\vec{M})$ between {\it the direction of the current and
 the magnetization}.  In single-domain structures, the angle $\theta$
 is tuned with the applied magnetic field which modifies the
 magnetization direction.  In contrast to GMR ($\uparrow \,
 \downarrow$ relaxation), AMR is a bulk effect that necessarily
 involves at least one anisotropic relaxation channel $\alpha
 \rightarrow \gamma (\theta)$ which is controlled by the direction of
 the magnetization (and is hence related to spin-orbit coupling)
 \cite{Potter}.  Although generated by spin-dependent electronic
 relaxations, the $\alpha \rightarrow \gamma (\theta)$ relaxation
 channel does not necessarily involve spin-flip scattering.  It is
 generally assumed that {\it the relaxation from the isotropic $s$
 minority channel $\alpha = s \downarrow$ to the anisotropic $d$
 minority channels $\gamma = d \downarrow$ is the main contribution to
 AMR in $3d$ ferromagnets } \cite{Mott,Potter0,Potter,Note2,cond-mat}. 
 In the {\it normal metal} (here normal means with no d band effect),
 the conductivity of the (minority) $d$ channel is vanishing, so that
 $\beta_{sd}^{N}=1$.  The out of equilibrium magnetoresistance is then
 a function of $\theta(\vec{M})$ defined by:

\begin{equation}
R^{N-F}_{AMR}(\theta) = \frac{1}{2} \frac{\left ( 1-\beta_{sd}(\theta) \right )^{2}}{
\sqrt{eL_{sd}(\theta) \, \sigma_{t}(\theta)(1-\beta_{sd}¥^{2}(\theta))}} 
\label{RAMR}
\end{equation}

where $\beta_{sd} (\theta)$ is the conductivity asymmetry
corresponding the AMR relaxation channels; $\sigma_{\alpha}(\theta)=
\sigma_{t}(\theta)(1+\beta_{sd}(\theta))/2$ and
$\sigma_{\gamma}(\theta)= \sigma_{t}(\theta)(1-\beta_{sd}(\theta))/2$ in 
the ferromagnet. 
In terms of diffusion length and normalized to the bulk AMR 
$R_{0}(\theta)$, the Ne-AMR writes :

\begin{equation}
	\frac{R^{N-F}_{AMR}(\theta) }{R_{0}(\theta)}=
	\left ( \frac{1-\beta(\theta)}{1+\beta(\theta)} \right ) \frac{l_{diff}(\theta)}{ 
	\Lambda}¥
	\end{equation}

However, the contribution of $R^{N-F}_{AMR}(\theta)$ is difficult to
measure because $l_{diff}$ is expected to be small (nanometric or
below), and the direct bulk contribution of the AMR dominates in usual
configurations (see however references \cite{Jedema,George} for a
possible contribution in $F_{1}¥/N/F_{2}$ devices).

\section{Out-of-equilibrium magnetothermopower}

Since, in metallic structures, the heat transfer is carried by the
conduction electrons, it is possible to study the electronic transport
coefficients by performing thermoelectric (TEP) measurements while
applying a temperature gradient to the sample.  TEP is usually
characterized through the bulk Seebeck coefficients, while imposing a
temperature gradient under zero electric current (open circuit).  In the
same manner as for GMR, TEP is composed of a bulk contribution and an
out-of-equilibrium contribution due to the interfaces (see next 
sub-section).  Surprisingly,
anisotropic MTEP in bulk ferromagnets has not been reported although
extensive investigations about TEP had been performed on Ni, Fe and Co
based materials since the work of Mott \cite{RqueTEP}. Thus, a vanishing 
bulk MTEP can be expected, that would favor the measurements of 
out-of-equilibrium interface MTEP.
Previous
investigations about the interface contribution to the
magneto-thermoelectric power (MTEP) have been performed exclusively in GMR 
structures, with typical sizes of the magnetic layers below the
spin-diffusion length (spin-valve structures)
\cite{MTEP,piraux,Shi,Shi2,Tsymbal,Gravier,Gravier2}.  In this very
case, the experimental results show that the spin-dependent
thermopower is nearly proportional to the GMR. As will be
shown below, the situation is similar in the case of single
ferromagnetic layers exhibiting AMR. 

In the following, the 
temperature gradient is assumed to be uniform : $ \nabla T = \frac{\Delta 
T}{\Lambda}$, where $\Lambda$ is the length of the wire, and $\Delta T $
is the temperature difference between the two terminals. This simplifying 
assumption allows us to recover the 
diffusion equation, Eq. (\ref{DiffEq}). The Onsager 
relations follow, by adding the heat flows $J^{Q}_{\alpha \gamma}¥$ of the two 
channels:

\begin{equation}
\begin{array} {lllll}
J_{\alpha} = -\frac{\sigma_{\alpha }}{e} \frac{\partial 
\mu_{\alpha}}{\partial z} + S_{\alpha} \sigma_{\alpha} \frac{\partial 
T}{\partial z}\\ 
J_{\gamma} = -\frac{\sigma_{\gamma}}{e} \frac{\partial \mu_{\gamma}}{\partial 
z} + S_{\gamma} \sigma_{\gamma} \frac{\partial 
T}{\partial z}\\ 
J^{Q}_{\alpha} = \lambda_{\alpha}¥ \frac{\partial 
T}{\partial z} -\pi_{\alpha}¥ \frac{\partial 
\mu_{\alpha}}{\partial z}\\ 
J^{Q}_{\gamma} = \lambda_{\gamma}¥ \frac{\partial 
T}{\partial z} -\pi_{\gamma}¥ \frac{\partial \mu_{\gamma}}{\partial z}\\ 
\dot{\Psi}_{\alpha \gamma} = L¥ \left ( \mu_{\alpha}-\mu_{\gamma} \right )
\end{array}
\label{OnsagerTEP}
\end{equation}

where $S_{i}$, $\lambda_{i}$, and $\pi_{i}$, $i=\{\alpha,\gamma \}$,
are respectively the Seebeck, the Fourier, and
the Pelletier coefficients of each channel.

Hereafter, we will not study the channel dependent heat flow
$J^{Q}_{\alpha \gamma}¥$.  The thermopower is deduced from Eqs
(\ref{OnsagerTEP}) following step-by-step the method developed in the
previous section, and incorporating the condition $J_{t}=0$.  In the
bulk metal, the local equilibrium condition leads to the relation:

\begin{equation}
J_{t}(\infty)= -\sigma_{t} \, \frac{\partial 
\Phi}{\partial z}(\infty) \, + \, S_{t} \sigma_{t} \frac{\Delta 
T}{\Lambda} = 0 
\end{equation}

which yields,

\begin{equation}
\frac{\partial \Phi}{\partial 
z}(\infty) = S_{t} \frac{\Delta T}{\Lambda}
\label{CurrentTEP}
\end{equation}

Where 

\begin{equation}
	S_{t}= \frac{\sigma_{\alpha}¥S_{\alpha} + 
\sigma_{\gamma}¥S_{\gamma}}{\sigma_{\alpha}+\sigma_{\gamma}¥}
\label{SRef0}
\end{equation}
is the {\it
the reference thermopower} corresponding to the bulk, or the equilibrium
TEP.
The effective current (analogous to the total current in the GMR 
calculation)
$J_{eff} = -S_{t} \sigma_{t} \frac{\Delta T}{\Lambda}$ is 
{\it different} in both
sides of the junction (like the conductivity, $\sigma_{t}$,
the Seebeck coefficient, $S_{t}$, is discontinuous at the interface). 

From Eqs. (\ref{OnsagerTEP}) and (\ref{currentcon}), the continuity 
of the currents $J^{I}_{\alpha}(0)=J^{II}_{\alpha}(0)$ leads to 
the following chemical-potential splitting at the interface: 

\begin{equation}
\Delta \mu (0)= \left ( \sigma^{I}_{\alpha} \, (S^{I}_{\alpha} - 
S^{I}_{t} ) \, - \, 
 \sigma^{II}_{\alpha} \, ( S^{II}_{\alpha} - 
 S^{II}_{t}) \right ) \,  \left ( \sqrt{\frac{\sigma^{I}_{\alpha} 
\sigma^{I}_{\gamma}eL^{I}}{\sigma^{I}_{t}}} + \sqrt{\frac{\sigma^{II}_{\alpha} 
\sigma^{II}_{\gamma}eL^{II}}{\sigma^{II}_{t}}} \right )^{-1} 
\, e \frac{\Delta T}{\Lambda}¥
\label{Deltamu}
\end{equation}

The chemical-potential splitting, $\Delta \mu (0)$, is analogous to that
calculated in Sec.  II, Eq.  (\ref{DeltaMu0}) for the GMR, after
introducing the effective current $J_{eff} = -S_{t} \sigma_{t}
\frac{\Delta T}{\Lambda}$ :

\begin{equation}
\Delta \mu (0)= e \left ( J^{I}_{eff}¥\frac{\sigma^{I}_{\alpha} - 
\sigma^{I}_{\gamma}}{\sigma^{I}_{t}} 
\frac{\sigma^{I}_{\alpha}¥}{\sigma^{I}_{t}} \frac{\sigma^{I}_{\gamma}¥}{\sigma^{I}_{t}}
- J^{II}_{eff}¥\frac{\sigma^{II}_{\alpha} - 
\sigma^{II}_{\gamma}}{\sigma^{II}_{t}} 
\frac{\sigma^{II}_{\alpha}¥}{\sigma^{II}_{t}} \frac{\sigma^{II}_{\gamma}¥}{\sigma^{II}_{t}} \right )  
\left ( \sqrt{\frac{\sigma^{I}_{\alpha} 
\sigma^{I}_{\gamma}eL^{I}}{\sigma^{I}_{t}}} + \sqrt{\frac{\sigma^{II}_{\alpha} 
\sigma^{II}_{\gamma}eL^{II}}{\sigma^{II}_{t}}} \right )^{-1} 
\label{DeltamuBIS}
\end{equation}

Here again (see Eq.  (\ref{Res})),
the out-of-equilibrium thermopower $\Sigma ^{ne}$ can be defined from the
reference corresponding to local equilibrium condition, $\Delta \mu_{eq}¥ = 
0$ and
$J_{\alpha}=J_{\gamma}=0$:

\begin{equation}
\Sigma^{ne} \frac{\Delta T}{\Lambda} = \frac{1}{e} \int_{A}^{B} \left ( 
\frac{\partial \mu_{\alpha}}{\partial z} - e S_{t} \frac{\partial 
T}{ \partial z} \right ) dz= \frac{1}{e} \int_{A}^{B} \left ( 
\frac{\partial \mu_{\alpha}}{\partial z} - e \frac{\partial 
\Phi}{ \partial z} \right ) dz 
\label{RefTEP}
\end{equation}

where $A$ (resp.  $B$) is located in the layer I (II), at a distance
$\Lambda^{I}$ ($\Lambda^{II}$), far enough from the interface
(inside the bulk).  This is the same expression as that calculated for
the out-of-equilibrium resistance in Eq.  (\ref{Result}). We obtain

\begin{equation}
\Sigma^{ne} \frac{\Delta T}{\Lambda} = -\left ( 
\frac{\sigma^{I}_{\alpha} - \sigma^{I}_{\gamma}}{ \sigma^{I}_{t}} -
\frac{\sigma^{II}_{\alpha} - \sigma^{II}_{\gamma}}{ \sigma^{II}_{t}} 
\right ) 
\frac{\Delta \mu(0)}{2e}
\label{MTEP}
\end{equation}

Making use of Eq.  (\ref{Deltamu})
we deduce the out-of-equilibrium TEP :

\begin{eqnarray}
\Sigma^{ne} = - \frac{1}{2} \left ( 
\frac{\sigma^{I}_{\alpha} - \sigma^{I}_{\gamma}}{\sigma^{I}_{t}} -
\frac{\sigma^{II}_{\alpha} - \sigma^{II}_{\gamma}}{\sigma^{II}_{t}} 
\right ) 
\left ( \frac{\sigma^{I}_{\alpha} \sigma^{I}_{\gamma}}{\sigma^{I}_{t}} 
\, (S^{I}_{\alpha} - S^{I}_{\gamma} ) \, - \, 
 \frac{\sigma^{II}_{\alpha} \sigma^{II}_{\gamma}}{\sigma^{II}_{t}} 
 \, ( S^{II}_{\alpha} - S^{II}_{\gamma}) \right ) \nonumber  \\
\left ( \sqrt{\frac{\sigma^{I}_{\alpha} 
\sigma^{I}_{\gamma}eL^{I}}{\sigma^{I}_{t}}} + \sqrt{\frac{\sigma^{II}_{\alpha} 
\sigma^{II}_{\gamma}eL^{II}}{\sigma^{II}_{t}}}\right )^{-1} 
\label{ResultMTEP}
\end{eqnarray}

Let us define the parameters, $\mathcal
S_{+}=(S_{\alpha}+S_{\gamma})/2$ and $ \mathcal
S_{-}=(S_{\alpha}-S_{\gamma})/2$.  We see that $S_{t}= \frac{1}{2}
\left ( (1+\beta) S_{\alpha} + (1-\beta) S_{\gamma} \right ) $, so 
that the 
overall
Seebeck coefficient rewrites:
 $$S_{t} = \mathcal S_{+} + \beta \mathcal S_{-}$$
The out of 
equilibrium interface thermopower takes the form: 

\begin{equation}
\Sigma^{ne}  = -(\beta^{I}-\beta^{II})
\frac{\sigma_{t}^{I} \left (1- (\beta^{I})^{2}¥ \right ) \,  
\mathcal S_{-}^{I} - 
\sigma_{t}^{II} \left (1-(\beta^{II})^{2} \right ) \mathcal S_{-}^{II}
}{\sqrt{eL^{I}\sigma_{t}^{I}(1-(\beta^{I})^{2})}+ 
\sqrt{eL^{II}\sigma_{t}^{II}(1-(\beta^{II})^{2})}}
\label{Sbeta}
\end{equation}

This is the general expression of the out-of-equilibrium MTEP. In the 
following, it will be expressed in terms of 
transport-coefficient asymmetry $\beta$.  
It is possible to investigate further this
relation by using the microscopic Mott's relation (valid for a
spherical energy band and assuming a local thermal equilibrium) 
\cite{Mott}:

\begin{equation}
S_{\alpha \gamma}¥=
\frac{a}{\sigma_{\alpha \gamma}¥} \left ( \frac{\partial
\sigma_{\alpha \gamma}}{\partial \epsilon} \right
)_{\epsilon_{F}}
\end{equation}

 where $a= \frac{\pi^{2} k_{B}^{2}T}{3e}$, $\epsilon$ is the electron
 energy, and $\epsilon_{F}$ is the Fermi energy.
 
 \begin{eqnarray}
\mathcal S_{+} =S_{t} - a \frac{ \beta \beta'}{1-\beta^{2}¥} \nonumber\\
\mathcal S_{-}¥ =  a \frac{\beta'}{1-\beta^{2}} 
\label{SRef}
\end{eqnarray}

and
\begin{equation}
S_{t}=\frac{a}{\sigma_{t}} \left ( \frac{\partial
\sigma_{t}}{\partial \epsilon} \right )_{\epsilon_{F}}
\label{Stotal}
\end{equation}

is the {\it the reference thermopower} defined in Eq.  (\ref{SRef0}),
and $\beta' = \frac{\partial \beta}{\partial\epsilon})_{\epsilon_{F}}$
is the derivative of the asymmetry conductivity coefficient $\beta$
taken at the Fermi level. Eq. (\ref{Sbeta}) rewrites : 
 
 \begin{equation}
\Sigma^{ne}  = 
-\frac{a ¥(\beta^{I}-\beta^{II}) \left ( \sigma_{t}^{I} \, 
\beta'^{I} - 
\sigma_{t}^{II} \, \beta'^{II} \right )
}{\sqrt{eL^{I}\sigma_{t}^{I}(1-(\beta^{I})^{2})}+ 
\sqrt{eL^{II}\sigma_{t}^{II}(1-(\beta^{II})^{2})}}
\label{Sbeta2}
\end{equation}

\subsection{Magnetothermopower corresponding to GMR and NeAMR}

In the case of spin-valve structures (i.e. junctions consisting of 
layers with parallel or
antiparallel magnetization), and considering identical ferromagnetic 
layers, we have $\beta_{s} =
\beta^{I} =-\beta^{II}$ and also $\beta'_{s} =
\beta'^{I} =-\beta'^{II}$ :

\begin{equation}
\Sigma_{GMR}^{\uparrow \downarrow} = 
- 2 a \sigma_{t} ¥\left ( \frac{\beta'}{\beta} \right ) \, R_{GMR}^{\uparrow \downarrow}
\label{SGMRGen}
\end{equation}

 As discussed in Ref. \cite{Shi}, the MTEP associated to GMR vanishes if the
 parameter $\beta' $ is zero, i.e. if the conductivity asymmetry is
 not energy dependent. The proportionality between $R^{GMR}/R_{0}$ and
 $\Sigma^{GMR}/(\Lambda S_{t})$ was observed experimentally \cite{Gravier,Gravier2,Shi,MTEP} 
 and the proportionality factor 
 $\mathcal P_{GMR}¥= - \frac{2 a}{S_{t}}¥\frac{\beta'}{\beta}$ was 
 found to be of the 
 order of one to ten in usual experimental conditions.

Besides, the out-of-equilibrium contribution due to the AMR
in a Normal/Ferromagnetic junction is deduced by taking into account
the relevant s-d relaxation channels: $\beta^{N}_{sd}=1$ 
(Sec. III.B) and $\beta'^{N}_{sd} = 0$ :

\begin{equation}
\Sigma^{N-F}_{AMR} = 2 a \sigma^{F}(\theta) \, \left ( 
\frac{\beta'(\theta)}{1-\beta(\theta)} \right )
R^{N-F}_{AMR}(\theta)
\label{SAMR}
\end{equation}

The expression $\Sigma^{ne}_{AMR}/S_{t} = \mathcal P_{AMR}
(R^{N-F}/R_{0})$ (where $R_{0} = \sigma_{t}(\theta)/ \Lambda$)
shows that a simple relation simmilar to that of GMR reltates the
NeAMR and MTEP. The proportionality factor $\mathcal P_{AMR}¥= \frac{2
a}{S_{t}¥} \frac{\beta'}{1-\beta}$ (refer to AMR/MTEP ratio in the
next section) can be measured providing that the NeAMR, described in
Sect.  III.B., Eq.  (\ref{RAMR}), is measured independently (e.g. with
the configuration proposed in references \cite{Jedema,George}).  The
relevance of the picture proposed above, which is based on the
differentiation between two well-separated relaxation channels
(spin-flip or s-d scattering) can now be compared to experimental
facts.

\subsection{Measuring MTEP}

It is important to point out that the measurements of interface TEP 
necessarily 
involve the measurement of the TEP of the bulk materials
contacted to the voltmeter through reference wires (see Fig.  1).  In our
experiments, a temperature difference $\Delta T = T_{B} - T_{A}$ is
maintained between the extremities A and B of the junction (located at
the J point), whereas the voltmeter with the terminals of the reference
wires are maintained at temperature $ T_{0}$.  Referring to the
TEP of the reference contact as $S_{t}$, the total voltage difference measured in
the open circuit consists of the bulk TEP and an interface TEP:

\begin{equation}
V_{TEP}¥ = \Delta T \left ( \frac{(AJ) S^{I}_{t} + (JB) S^{II}_{t}}{AB} 
 - S_{r} \right) + \Sigma^{ne} \left ( \frac{\partial T}{\partial z} \right )_{J}¥
\label{TotalMTEP}
\end{equation}

As already pointed out, and according to the literature, the bulk
term appears to be independent on the magnetic configuration (i.e. 
independent on $ \theta$).  Such a
situation occurs under the following weakly restrictive condition:
$\sigma_{t}(\epsilon, \theta) = g(\theta) \sigma_{t}(\epsilon)$ (see 
Eq. (\ref{Stotal})), where
$g(\theta)$ is any function accounting for the conductivity anisotropy. 
In contrast, the out-of-equilibrium term is still $\theta$ dependent 
through the parameter  $\beta(\theta)$, or $l_{sd}(\theta$).
In consequence, we expect that a MTEP contribution can be measured as a
function of the external magnetic field, and that this MTEP is produced by the
out-of-equilibrium interface term only.  On the other hand, the amplitude of the
non-equilibrium interface effect depends on the amplitude of the
temperature gradient at the junction $\left ( \frac{\partial
T}{\partial z} \right )_{J}$.  The effect is then larger in case of
a non-homogeneous temperature gradient, if the junction is placed in a region where
there is a sharp temperature variation, i.e. near the interface with
the heat source or cryostat.  In contrast, if the junction is placed
far away from the interface with the heat source or cryostat, the
effect is expected to be smaller.

As for AMR, the $\theta$ dependence of the TEP (the MTEP) is defined
as the ratio: 

\begin{equation}
\frac{\Delta V}{V} =	\frac{Max\{V(\theta)\}-Min\{V(\theta)
\}}{Min\{V(\theta)\}}
\end{equation}
In the next 
section, the quantity
$V(\theta)$ is measured as a function of the amplitude and direction of
the applied magnetic field $\vec H$.

\begin{figure}
   \begin{center}
   \begin{tabular}{c}
   \includegraphics[height=6cm]{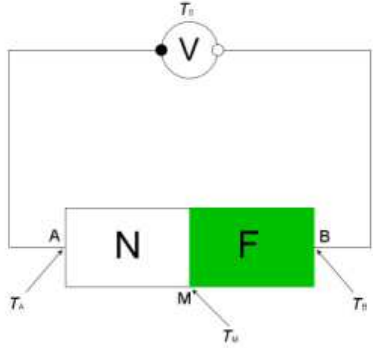}
   \end{tabular}
   \end{center}
   \caption[Potentiel] 
{ \label{fig:SetUp} The structure consists of two metallic layers of
length AJ and JB with a typical temperature gradiant $\Delta T / AB $. 
It is contacted through two reference wires connected to a voltmeter
at temperature $T_{0}$}
   \end{figure}

\section{Experiments}

As already mentioned, the nearly linear relation between the GMR
($\Delta R/R$) and the corresponding MTEP ($\Delta V/V$) has been
observed in various spin-valve systems
\cite{MTEP,piraux,Shi,Shi2,Tsymbal,Gravier,Gravier2}.  The GMR/MTEP ratio
is of the order of one to ten in GMR samples consisting of about 150
electrodeposited Co/Cu bilayers where both the GMR and the MTEP are of
the order of 10 \% \cite{Gravier}.  The present study focuses on MTEP
in single Ni nanowires by pointing out the role of the contacts.  The
results presented hereafter have been measured near room temperature. 
All nanowires contain two contacts N/F and F/N, and a
bulk ferromagnetic (F) region.  The results presented in Sec. IV
predict that an anisotropic out-of-equilibrium interface
magnetoresistance, and corresponding MTEP, should be present at the 
junctions.

This experimental section is composed as follows.  The samples are
described in subsection A. The magnetic configurations of the nanowire
are discussed in subsection B on the basis of recent AMR measurements
and of previous reports.  Subsection C reports on the anisotropic
nature of the measured MTEP. Section D evidences that the measured MTEP is
an interface effect.  Subsection E describes the magnetic
configurations of the Ni contact that allow the MTEP profiles to be
understood.

\subsection{Samples}

The samples are prepared by electrodeposition in porous polycarbonate
track-etched membranes.  This technique has been used extensively in
order to study the micromagnetic configurations inside the wires
\cite{Travis,Fert,IEEE,Schoen,Yvan,PRL,Meyer}.  The pores are
6-micrometer length and 40 to 25-nm diameter.  A gold layer is
deposited on the bottom and top of the membrane and fixed to the
electrode.  By applying the potential in the electrolytic bath, the Ni
nucleates at the bottom of the pores, grows through the membrane and
reaches the top Au layer.  Then, a single nanowire can be contacted inside
the electrolytic bath, by controlling the potential between the two
sides of the membrane during the electrodeposition and stopping the
process when the potential drops to zero \cite{IEEE}.  The single
contact can be performed either with the same material as that of the
wire (Ni) or with a different material (for instance non-ferromagnetic
like Cu or Au), by changing the electrolytic bath before performing
the contact (see Fig.  2).  The contact has the shape of a mushroom on
top of the membrane \cite{IEEE,Yvan,Schoen}.

The electrodeposited Ni nanowire consists of nanometric
nanocrystallites with random orientations: the magnetocrystalline
anisotropy is averaged out at the nanometer scale
\cite{Meyer,Yvan,Derek,PRL}.  Only a strong uniaxial shape anisotropy
remains present (anisotropy field $H_{a}= 2 \pi M_{S} \approx 0.6$ T, where
$ M_{s}$ is the magnetization at saturation).  It has been
shown that the Ni nanowires are uniformly magnetized for all stable
states \cite{PRL,Yvan}.  Furthermore, due to the high aspect ratio, the
spatial distribution of the current density $\vec{J}$ is well defined along
the wire axis: the angle $(\vec{J}, \vec{M}) $ between the 
current and the magnetization $\vec{M}$ coincides with the
angle $\theta$ of the magnetization of the wire (see Fig.  2).

It is expected that a ferromagnetic contact localized on the top of
the membrane (the Ni mushroom) changes the interface properties for
two reasons: due to the non uniform spin-polarized current density
\cite{Otani}, and due to the presence of specific magnetic
configurations that do not exist inside the wire. Note that the problem related to the
spin-accumulation and GMR generated by magnetic domain walls has been studied
in detail in such electrodeposited samples \cite{DWS}.  {\it The
conditions that are necessary to obtain a GMR-like contribution, 
the presence of a highly-constrained magnetic domain wall, are not
fullfilled} in the present case.  Here we report on a comparative study
between samples with different contacts for a
significant number of samples (a few tens).

\begin{figure}
   \begin{center}
   \begin{tabular}{c}
   \includegraphics[height=9cm]{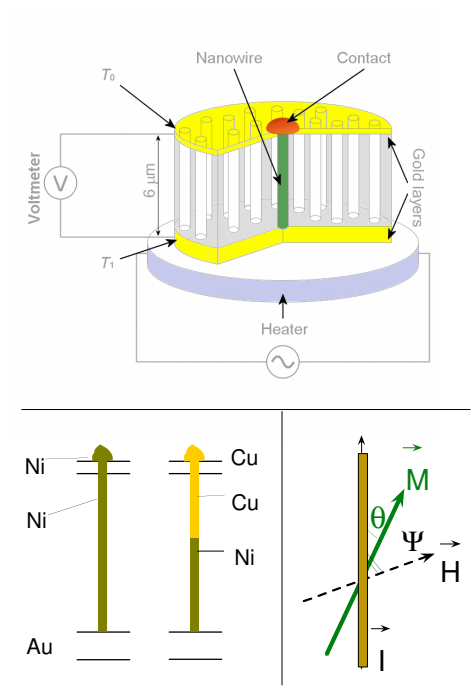}
   \end{tabular}
   \end{center}
   \caption[Potentiel] 
{ \label{fig:SetUp} Geometry and contacts of the two kinds of single 
contacted nanowires.  The heat resistance at the bottom is driven by 
an AC voltage generator at
frequency f.}
   \end{figure}

\subsection{Magnetic characterization through AMR}

Due to the uniform magnetization and to the homogeneous current density,
the magnetic field dependence of the AMR is directly linked to the
magnetic hysteresis loop of the Ni nanowire.  A
quadratic dependence is observed \cite{Potter}:

\begin{equation}
R(\theta)=R_{0}+ \Delta R_{AMR} \, \, cos^{2}( \theta )
\label{AMRexp}
\end{equation}

The magnetoresistance (Fig.  3) is measured with an external
magnetic field applied at a given angle $\Psi$ with respect to the wire axis. 
Except for some few samples were domain walls can be observed (not
shown), the hysteresis loop corresponds to a uniform rotation of the
magnetization with a precision of two to three percents
\cite{PRL,Yvan,Marcel}.  The magnetic configurations are
described by the well-known profile (see e.g. the 
Stoner-Wohlfarth model) \cite{Aharoni}. At large
angles ($\Psi \approx 90 (deg)$), the magnetization states follow a
reversible rotation from the wire axis $\theta = 0$
to the angle of the external field $\Psi $ while increasing the
magnetic field from zero to the saturation field (see Fig.  2):
intermediate states ($\theta \in [0,90]$) are stable and correspond to
the profile of the AMR curve (Fig.  3).  In contrast, for small
angles (around $\Psi \approx 10$ deg) the magnetoresistance profile as
a function of the applied field (Fig.  3) is flat because the
magnetization is pinned along the wire axis : there are no stable
positions between $\theta \approx 10$ and $\theta \approx 170$ deg. 
There is no fundamental change if contacting the nanowire with Cu
or Au \cite{Marcel}.

\begin{figure}
   \begin{center}
   \begin{tabular}{c}
   \includegraphics[height=10cm]{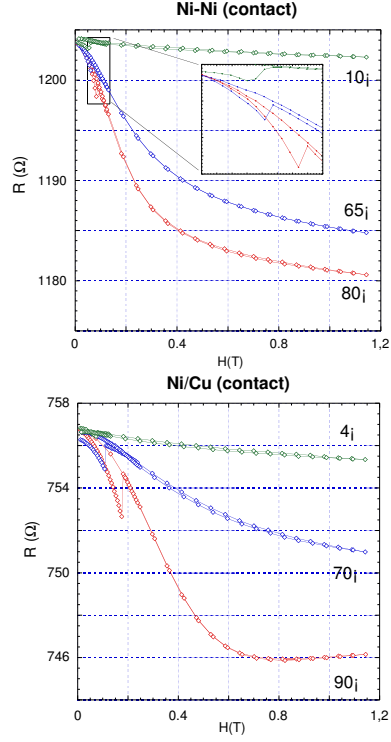}
   \end{tabular}
   \end{center}
   \caption[Potentiel] 
{ \label{fig:AMR} The AMR is plotted at different angles of the 
external field: (a) Ni wire contacted with Ni; (b) Ni wire contacted with 
Cu.}
   \end{figure}

  \subsection{MTEP is anisotropic}
  
The thermoelectric measurements are performed with a compact resistive heater
(5 Ohms), placed on the bottom of the membrane and contacted to a
voltage generator of 5 to 7 Volts (Fig.  2).  A sine wave of frequency
of the order of f=0.05 Hz is injected in the heater.  At this
frequency, a stationary thermal regime is reached, and the output
thermopower signal is detected at 2f = 0.1Hz.  The amplitude of the 2f
signal gives the TEP $\Sigma^{ne} \Delta T/ \Lambda$.  With our
experimental configuration, the amplitude of the TEP ranges between
five to fifty $\mu V$, which corresponds to $\Delta T \approx 1 K$,
with $S_{T}^{Ni} \approx -13$ $\mu$ V/K and $S^{Cu}_{t} \approx 1.8
\mu $ V/K, so that the temperature gradient is $\frac{\Delta
T}{\Lambda} \approx 3 \, 10^{5} K/m$. These values are close to
that measured in electrodeposited Co/Cu/Co multilayered spin-valves
\cite{Gravier,Gravier2}.

A MTEP signal is obtained by measuring the voltage at zero current, as
a function of the applied field.  The MTEP signal does not originate 
directly from the magnetic field, but is related to the
ferromagnetic configurations: the {\it anisotropic nature of the MTEP} is
observed in Fig.  4, by measuring the TEP voltage as a function of the angle
of the applied saturation field (at saturation field, the
magnetization aligns with the field : $\theta = \Phi$).  The anisotropic
MTEP, with a $\Delta V / V$ variation of about 13 \%, can be compared
to the corresponding AMR (1.3 \% amplitude, fitted with a $cos^{2} 
\theta$ law) in Fig.  4.  The MTEP profile is not very regular, and varies slighly 
 from one sample to the other. 

    \begin{figure}
   \begin{center}
   \begin{tabular}{c}
   \includegraphics[height=8cm]{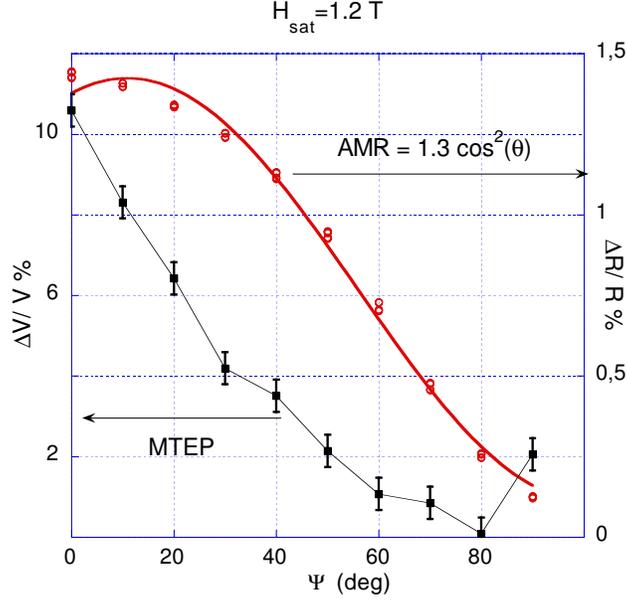}
   \end{tabular}
   \end{center}
   \caption[Potentiel] 
{ \label{fig:angles} Comparison of magnetothermopower (left) and 
AMR (right) for a single Ni nanowire 
with a Ni contact measured as a function of the angle of the external 
field with a saturating field ($\theta = \Psi$) of 1.2 T.}
   \end{figure}

The typical MTEP signal of Ni nanowires contacted with Ni, measured as
a function of the external field, is shown in Fig.  5 for the sample
characterized in Fig.  3 (a).  A variation larger than that of the AMR
signal is seen (depending on the samples, the MTEP amplitude ranges
from about $\Delta
V/V = 3 $ \% up to 30 \%) and is of the same order that the MTEP
produced in GMR devices composed of 150 bilayers \cite{Gravier}.  The
overall shape is surprising, since the profile as a function of the
external field $\vec{H}_{ext}$ at small angles $\Psi$ shows the
maximum variation (while the magnetization is fixed along the wire
axis), and inversely, the profile at large angle $\Psi$ is
approximately flat (while the magnetization rotates from zero to 90
deg).  Note that the MTEP minimum at small angles corresponds to the
zone of switching field (see Fig.  3), and that the high-field profile
shows an approach to saturation corresponding to the anisotropy field
of the wire.  Such curves are systematically observed on all measured
samples with small diameters (about 15 samples of diameter about 40 nm) \cite{LastMeas}.

\subsection{MTEP is not a bulk effect}

The MTEP profile is not a
function of the angle $\theta$ between the magnetization of the Ni
nanowire and the wire axis and the variations observed should be related 
to another parameter. The most likely hypothesis is that the variations 
are produced by the magnetization states confined at the interface close to the Ni 
contact. {\it In contrast to the AMR which is a bulk 
effect, the MTEP appears as an interface out-of-equilibrium process}.

\begin{figure}
   \begin{center}
   \begin{tabular}{c}
   \includegraphics[height=10cm]{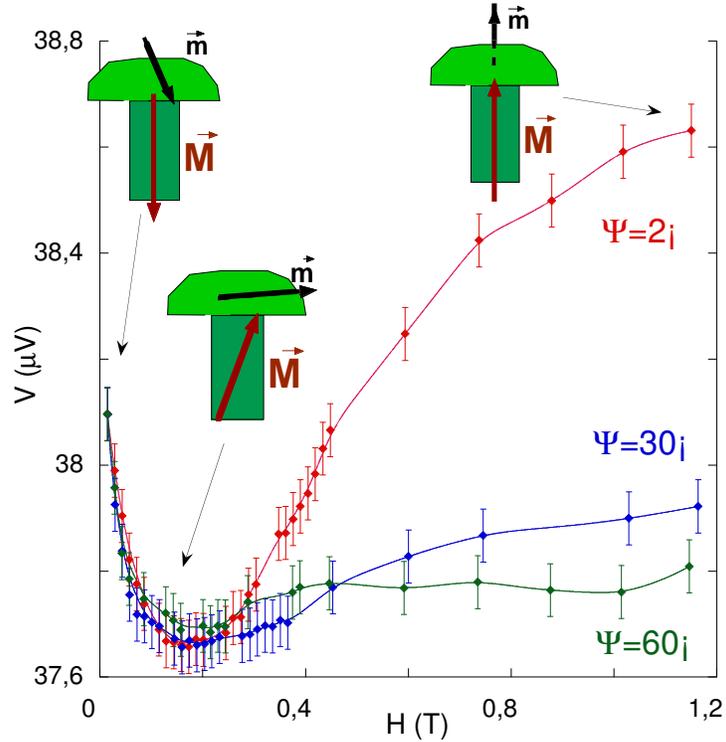}
   \end{tabular}
   \end{center}
   \caption[Potentiel] 
{ \label{fig:3} Thermoelectric power as a function of the external 
field in Ni nanowires contacted with Ni for different directions of the 
external field. The magnetic configurations of the Ni contact
are represented with arrows for $\Psi \approx 2$ deg.}
   \end{figure}

This hypothesis can easily be checked by comparing the Ni nanowires
contacted with Ni to those contacted with Cu or Au (See Fig. 2). In these
last samples, the ferromagnetic/normal interfaces are located inside the
nanowire where electric current, temperature gradient, and magnetization are 
homogeneous. We observe that the MTEP signal
vanishes with Cu and Au contacts (the TEP measured as a function of
the angles $\Psi$ is constant).  The two curves measured as a function
of the applied field are compared in Fig. 6 (concerning the two samples
characterized in Fig.  3), for $\Psi=0$. These measurements first confirm that
the effect is due to the interface, and second, that the role played
by the Ni contact is essential for the
observation of MTEP processes.  Note that a similar role of the
Ni contact has been observed in experiments of spin-injection induced
magnetization switching \cite{Marcel}, were irreversible
magnetization reversal provoked by the current was observed with 
ferromagnetic
contacts, but not with Cu contacts.

   \begin{figure}
   \begin{center}
   \begin{tabular}{c}
   \includegraphics[height=10cm]{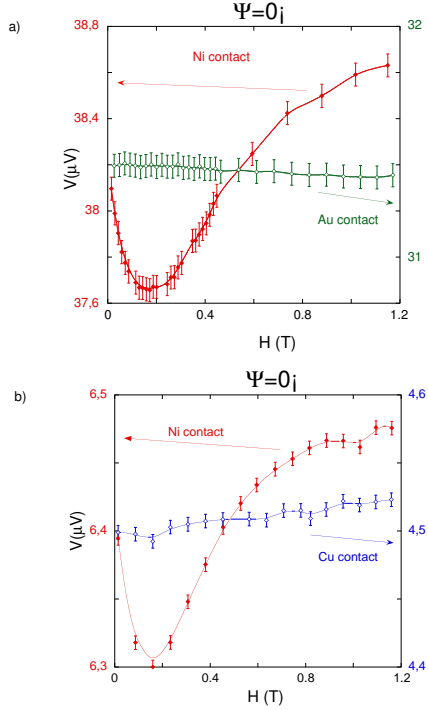}
   \end{tabular}
   \end{center}
   \caption[Potentiel] 
{ \label{fig:Contacts} Magnetothermopower of Ni nanowires with Ni, Au and
Cu contacts, measured with external field $\Psi \approx 0$ deg (a): Ni with
Ni contacts (left scale) compared with Au contact (right scale), MTEP
$\approx 0.3 \%$ for about $6 \mu $V TEP; (b) Ni with Ni contactss
compared with Cu contacts, MTEP $\approx 3.3 \%$, for TEP about 38
$\mu$V}
   \end{figure} 

  These observations corroborate the analysis performed in Sec.  IV. B
  where the amplitude of the effect is shown to be proportional to
  $\Sigma^{ne} \Delta T/AB$.  In the case of an interface localized
  near the mushroom, the ratio $l_{diff}/AB$ is very large.  In
  contrast, in the case of Cu or Au contacted wires, where the
  ferromagnetic/normal junction is localized inside the wire, the
  ratio $l_{diff}/AB$ is expected to be much smaller.

 \subsection{MTEP is related to the magnetic configurations of the Ni 
 contact}

 It is possible to relate the observed MTEP to the AMR if we consider
 that the relevant angle is the angle
 $\theta_{N/F}=(\vec{I}_{N/F},\vec{M}_{N/F}¥)$ between the local
 current $I$ and the magnetization $M$ at the nanoscopic scale near
 the $N/F$ interface.  With Cu and Au contacts, both the current
 density and the magnetization direction are well defined, and the
 angles coincide with that of the AMR:
 $\theta_{N/F}=(\vec{I}_{N/F}¥,\vec{M}_{N/F})= \theta$.  However, with
 the Ni contact, the interface is located near the Ni mushroom.  The
 direction of the current is no longer along the wire axis, and the
 magnetic configurations do not follow that measured with AMR inside
 the wire (see Fig.  5).  The relevant angle $\theta_{N/F}¥ $ is
 defined by the current direction in the Ni mushroom, probably near
 the plane of the contact (if the current were uniform).  The MTEP
 variations can then be reproduced assuming that the magnetization of
 the mushroom rotates following the total magnetic field $\vec{H}_{T}
 = \vec{H}_{a} + \vec{H} + \vec{H}_{perp}$ where $\vec{H}_{a}$ is the
 dipole field due to the wire (which is of the order of the shape
 anisotropy of the wire) and $\vec{H}$ is the applied field.  The
 field $\vec{H}_{perp}$ is the shape anisotropy of the mushroom.  It
 is produced by the dipole field of the mushroom, probably interacting
 with the other vicinity mushrooms in the plane of the membrane (it
 plays the role of the anisotropy field of a thin layer). Thus the 
 case of large and small angles have to be distinguished :  i) The
 application of the external fields at large angles fixes the
 magnetization of all mushrooms in the plane perpendicular to the wire
 axis so that the configuration with the magnetization of the mushroom along
 the wire axis is expected only near zero applied field where $H_{a}$
 dominates.  ii) In the case of an external magnetic field applied at
 small angles $\Psi \le $ 10 $^{o} $ ( see schemes of Fig.  5), the
 magnetization of the mushroom is along the wire axis for nearly zero
 field ($H_{a}$ dominates) and for saturation fields ($H$ dominates). 
 At intermediate fields, the magnetization of the wire switches to the
 opposite direction: a domain wall should be present between the wire and the
 mushroom.  The transverse field $\vec{H}_{perp}$ dominates.  The above
 scenario describes well the curves observed at different angles: the
 minima correspond to the MTEP with the magnetization of the mushroom
 perpendicular to the wire axis.  The maximal value of MTEP
 corresponds to the magnetization of the mushroom parallel to the wire
 axis.  The whole behavior is similar to that of AMR (see Fig.  4).
 
   \section{conclusion}

The well-known two-spin-channel model has been extended to the general
case of an interface between two layers in the relaxation time
approximation.  A general expression of the thermoelectric power is
derived.  Like giant magnetoresistance (GMR), a non-equilibrium
interface resistance contribution due to the anisotropic
magnetoresistance (AMR) is predicted in a ferromagnetic/normal
interface due to s-d interband relaxation.  The corresponding
magnetothermopower (MTEP) is derived, and is found to be proportional
to $ l_{diff} \left ( \frac{\partial T}{\partial z} \right )_{J}¥$
where $ l_{diff}$ is the relevant diffusion length, and
$\left ( \frac{\partial T}{\partial z} \right )_{J}$ is the temperature gradient at the
junction (see Fig.  1).  The MTEP associated to GMR is proportional to
the magnetoresistance with the proportionality coefficient $\mathcal
P_{GMR}¥ =- \frac{2 a}{S_{t}}¥\beta' / \beta$ and the MTEP associated
to AMR is proportional to the out of equilibrium AMR, with the
coefficient $\mathcal P_{AMR}¥ = \frac{2 a}{S_{t}} \beta' /
(1-\beta)$.  In the case of GMR, the experimental value of $\mathcal
P_{GMR}$ is close to one \cite{Gravier} (the MTEP is of the same
order as the GMR) for many junctions in series.

In complement to the experiments with multilayered systems (Co/Cu/Co)
\cite{Gravier}, measurements of MTEP in electrodeposited Ni nanowires
are presented.  This signal presents three striking features:
(i) A large MTEP signal of several $\mu V$ for about 1K temperature
variation is measured (3 to 30\% of the TEP);  (ii) This MTEP is
anisotropic;  (iii) The measured MTEP signal is produced by a local
magnetic configuration (at nanometric range) near the interface only. 
However, in contrast to transport experiments in GMR systems where
both the magnetoresistance and the magnetothermopower are measured,
the out-of-equilibrium AMR is not accessible in our two-point
measurements in Ni nanowires.  Accordingly, the interpretation of
anisotropic MTEP due to GMR (where $ MTEP \propto l_{sf}/AB$) produced
by magnetic inhomogeneities (i.e. domain-wall scattering effects) 
cannot be directly rule-out.  But the interpretation of domain wall TEP is
not realistic because DWS is very weak (below 0.1 \% if any, according
to previous studies \cite{DWS}) so that an important anisotropic MTEP
could be measured only with a huge proportionality coefficient ($\ge
100$), which is in contradiction with the known GMR
coefficient ($\mathcal P_{GMR} \approx 1$ for 150 junctions)
measured in GMR structures.

The results of this study hence show that, while GMR and associated
thermopower indicates spin-flip diffusion at the interface, the
observed interface anisotropic MTEP should indicate interband s-d
relaxation associated with ferromagnetism in Ni (where $MTEP \propto
l_{sd}/AB$).  The amplitude of the effect suggests that the
corresponding sd-diffusion length is sizable (e.g. of the order of the
spin-flip length $l_{sf}$¥). Within this framework, further experiments
allowing direct measurements of non-equilibrium AMR would probe and
clarify the role played by the two kinds of relaxation processes.

\section{Acknowledgement}
HJD thanks the D\'el\'egation G\'en\'erale pour l'Armement for
support.


\begin{references}
\bibitem{hybride} The case of up and down spin electrons is obvious. 
In the case of $s$ and $d$ channel, band hybridization was often invoked 
against this distinction. However, even when dealing with 
hybridized states, it is still possible to separate conduction 
channels with a dominant $s$ character from conduction channnels 
originating from states with a stronger $d$ contribution.
\bibitem{Constantes} In all these calculations, the channel parameters  
$\sigma_{\alpha \gamma}$, 
$\beta$, and $L$ are considered as constant in space. It can be shown 
that taking into account the gradient only introduces vanishingly 
small corrections.
\bibitem{LastMeas} Some nanowires with very high AMR (about 3 \%) and 
large diameter show 
a MTEP profil as a function of the magnetic field that follows the AMR 
profile: these samples have a very low contact resistance, and no 
mushroom effects.
\end{references}
\end{document}